\documentclass[11pt, a4paper]{article} 

\usepackage[ansinew]{inputenc}
\usepackage{amsmath, amssymb, graphics, amsthm}
\usepackage{epsfig}
\usepackage{color, xcolor}
\usepackage{fancyhdr} 
\usepackage{manfnt}
\usepackage[T1]{fontenc}

\usepackage[normalem]{ulem}

\newcommand{\RB}{\mathbb R_\text{Bohr}}

\newcommand{\ket}[1]{\left| #1 \right\rangle}

\newcommand{\twopoint}[1]{\left\langle #1 \right\rangle}

\newcommand{\vev}[1]{\left\langle #1 \right\rangle}

\newcommand{\braket}[2]{\left\langle \vphantom {#1 #2} #1 \hphantom{|} \right| \left. \vphantom {#1 #2} #2 \right\rangle}

\newcommand{\braopket}[3]{\left\langle \vphantom {#1 #2 #3} #1 \hphantom{|} \right| #2 \left| \hphantom{|} \vphantom {#1 #2 #3} #3 \right\rangle}

\makeatletter
\@addtoreset{equation}{section}
\makeatother

\newcommand{\be}{\begin{equation}}
\newcommand{\ee}{\end{equation}}
\newcommand{\ba}{\begin{eqnarray}}
\newcommand{\ea}{\end{eqnarray}}

\def\pb#1{\rlap{\lower1.5ex\hbox{$\longleftarrow$}}{#1}}
\def\dpb#1{\rlap{\lower1.5ex\hbox{$\Longleftarrow$}}{#1}}
\def\spb#1{\rlap{\lower1.0ex\hbox{$\leftarrow$}}{#1}}
\def\sdpb#1{\rlap{\lower1.0ex\hbox{$\Leftarrow$}}{#1}}

\DeclareMathOperator{\sign}{sign}

\title{{\sf State refinements and coarse graining in a full theory embedding of loop quantum cosmology}}
\author{
{\sf N. Bodendorfer}\thanks{{\sf 
norbert.bodendorfer@fuw.edu.pl}}\\
\\
{\sf  Faculty of Physics, University of Warsaw, Pasteura 5, 02-093, Warsaw, Poland}\\
}
\date{{\small\sf \today}}

\begin{document} 

\maketitle

{\sf

\begin{abstract}

Bridging between descriptions involving few large and many small quantum numbers is the main open problem in loop quantum gravity. In other words, one would like to be able to represent the same physical system in terms of a few ``coarse'' quantum numbers, while the effective dynamics at the coarse level should agree with the one induced by a description involving many small quantum numbers. Efforts to understand this relationship face the problem of the enormous computational complexity involved in evolving a generic state containing many quanta. In a cosmological context however, certain symmetry assumptions on the quantum states allow to simplify the problem. In this paper, we will show how quantum states describing a spatially flat homogeneous and isotropic universe can be refined and coarse grained. Invariance of the dynamics of the coarse observables is shown to require a certain scaling property (familiar from loop quantum cosmology) of the quantum states if no running of parameters is taken into account. The involved states are solutions to the Hamiltonian constraint when terms coming from spatial derivatives are neglected, i.e. one works in the approximation of non-interacting FRW patches. The technical means to arrive at this result are a version of loop quantum gravity based on variables inspired by loop quantum cosmology, as well as an exact solution to the quantum dynamics of loop quantum cosmology which extends to the full theory in the chosen approximation.

\end{abstract}

}

\section{Introduction}

Recent advances in loop quantum cosmology (LQC),  see \cite{AshtekarLoopQuantumCosmologyFrom} for a review, including windows for experimental predictions \cite{AgulloThePreInflationary, DeBlasPrimordialPowerSpectra} and the establishment of falsifiability \cite{BollietObservationalExclusionOf}, underline the necessity to understand the precise relation of full loop quantum gravity (LQG) with LQC. To this end, several approaches have been developed recently using a wide variety of technical means \cite{AlesciANewPerspective, AlesciLoopQuantumCosmology, GielenHomogeneousCosmologiesAs, BIII, BVI, OritiEmergentFriedmannDynamics, BeetleDiffeomorphismInvariantCosmological}. A central questions in all those approaches is the choice of quantum state describing a homogeneous and isotropic universe: while states involving few large quanta (or parameters) are often possible to handle, the dynamical equivalence to a description involving many small quanta usually remains obscure. A middle way is to consider states involving many quanta, but subject them to certain (ad hoc) symmetry assumptions, such as considering a collection of non-interacting quanta in the same state, e.g., as for condensates \cite{GielenHomogeneousCosmologiesAs}. 

In this paper, we will show that a recent proposal for a full theory embedding of LQC can be amended in this way, i.e. that the derivation of the quantum dynamics which previously used only a single vertex to describe the whole universe can also be formulated using an arbitrary number of vertices. 
A key assumption in the derivation is to neglect terms in the Hamiltonian constraint corresponding to spatial derivatives. For a certain choice of solutions where the wave functions at all vertices is the same, it is shown that the expectation value of the neglected terms vanishes, establishing self-consistency of the approximation. 
These results can be derived analytically, owing to an exact solution of LQC \cite{AshtekarRobustnessOfKey} in a suitable ordering which we import into the full theory.  

The findings of this paper thus establish that for a certain choice of quantisation variables and in a highly symmetric subsector of the theory, the issues of coarse graining and refinement can be explicitly addressed. 
The results however heavily rely on a certain choice of quantisation variables which allows to implement the $\bar \mu$-scheme \cite{AshtekarQuantumNatureOf} of loop quantum cosmology in the full theory. In particular, it is unclear whether they can be extended to Ashtekar-Barbero variables \cite{AshtekarNewVariablesFor, BarberoRealAshtekarVariables}, which are normally used in LQG, due to an obstruction in implementing the $\bar \mu$-scheme in a graph-preserving way for SU$(2)$ as a gauge group \cite{BIII, BVI}. Still, more elaborate constructions such as graph superpositions \cite{AlesciImprovedRegularizationFrom} or the group field theory dynamics \cite{GielenHomogeneousCosmologiesAs, OritiEmergentFriedmannDynamics} could help out.  \\

This paper is organised as follows:\\
In section \ref{sec:Recap}, we review the main ideas of \cite{BVI} and highlight the motivations entering the various construction steps. Next, we show in section \ref{sec:Refinements} how a natural set of state refinements does not change the coarse dynamics. We conclude in section \ref{sec:Conclusion}.

\section{A full theory embedding of LQC} \label{sec:Recap}

An important virtue of LQC is its simplicity owing to the large amount of symmetry, which allows to explicitly perform numerical and in special situations even analytical computations. Within full LQG, it would be strongly desirable to find similar simplifications, and ideally have the theory reduce to LQC under the same symmetry assumptions. Within the standard formulation of LQG in terms of Ashtekar-Barbero variables, it is unclear whether this goal can be achieved. However, it is possible if one changes the quantisation variables in such a way that a natural (symplectic) split between those relevant for a Friedmann-Robertson-Walker (FRW) universe, the volume and mean curvature, and the other variables occurs. We will review in the following how this can be done and refer to \cite{BVI} for details. 

\subsection{Classical reformulation}

We now present a brief review of the results of \cite{BVI}, focussing on the parts relevant for the current paper. 
We start with the Arnowitt-Deser-Misner (ADM) formulation of general relativity \cite{ArnowittTheDynamicsOf}, where phase space is coordinatised by the spatial metric $q_{ab}$ and its conjugate momentum $P^{ab}$. The Poisson brackets read
\be
	\{ q_{ab}(x), P^{cd}(y) \} = \delta^{(3)}(x,y) \delta_{(a}^c \delta_{b)}^d \text{.}
\ee
We restrict the spatial slice $\Sigma$ to be a 3-torus for simplicity. 
In order to describe a FRW universe, we are interested in the volume density $\alpha := \sqrt{q}$ as well as the canonically conjugate variable $P_\alpha = \frac{2 P^{ab} q_{ab}}{ 3 \sqrt{q}}$, resulting in $\{ \alpha(x), P_{\alpha}(y)\} =\delta^{(3)}(x,y)$. 
The motivation behind this choice is a direct link to the $b,\nu$-variables of loop quantum cosmology \cite{AshtekarRobustnessOfKey}, which are related as $b \propto P_\alpha$ and $\nu \propto \int d^3x \, \alpha$.
From the remaining diagonal components of the spatial metric and its momentum, we assemble further variables which Poisson commute with both $\alpha$ and $P_\alpha$, so that we can treat them separately in the quantum theory. 
These variables will measure the deviation of the metric and its momentum from the spatially flat, homogeneous and isotropic form $q_{ab} \propto \delta_{ab}$ and $P^{ab}  \propto  \delta^{ab}$. As an example, the choice $\beta = P^{xx}q_{xx} - P^{yy} q_{yy}$, $P_\beta = \frac{1}{2} \log \left( q_{yy}/q_{xx} \right)$ as one of the two additional canonical pairs is made in \cite{BVI}, the second one, constructed similarly, will be denoted by $\gamma$ and $P_\gamma$.
Other details of this procedure are irrelevant for the current paper and can be found in \cite{BVI}. The diagonal metric gauge for the spatial diffeomorphism constraint is imposed, removing the off-diagonal components of the spatial metric. The off-diagonal components of $P^{ab}$ are solved for by the spatial diffeomorphism constraint. While this is in general not possible analytically, it turns out not to be necessary for describing a FRW universe due to the following. 

Next, we have to discuss a set of constraints, denoted as ``reduction constraints'', which we are going to impose in the quantum theory. These constraints are derived classically as phase space functions which vanish in a FRW universe, in our case in a certain set of (standard cartesian) coordinates. The choice of these constraints is pure convenience and we in particular do not demand that the set of constraints is classically sufficient to impose homogeneity and isotropy. Rather, they are chosen in such a way that they can be easily implemented in the quantum theory, in particular they are first class, and remove certain unwanted terms. A suitable set of reduction constraints contains the generator of spatial diffeomorphisms acting on $\alpha$ and $P_\alpha$, 
as well as on additional matter fields present in the theory, here taken to be a scalar field $\phi$ and its conjugate momentum $P_\phi$:
\be
	C_a [N^a] = \int_\Sigma d^3x \left( P_\alpha \mathcal L_{\vec N} \alpha + P_\phi \mathcal L_{\vec N} \phi \right)  \text{.} \label{eq:DiffRedConstr}
\ee
Furthermore, the set of reduction constraints contains conditions on the other variables measuring the departure from $q_{ab} \propto \delta_{ab}$ and $P^{ab}  \propto  \delta^{ab}$. 
These can be easily imposed, since they amount to the vanishing of the additional gravitational variables $\beta$, $P_\beta$, $\gamma$, and $P_\gamma$.

\subsection{Quantisation}

Quantisation is achieved by standard techniques from loop quantum gravity, see \cite{ThiemannModernCanonicalQuantum}. Since $P_\alpha$ transforms as a scalar under spatial diffeomorphisms (generated by \eqref{eq:DiffRedConstr}) and $\alpha$ as a density, we can simply use the quantisation prescription for a scalar field as given in \cite{ThiemannKinematicalHilbertSpaces}. Cylindrical functions $\Psi$ depend on point holonomies $h^\rho_x := e^{i \rho P_\alpha}(x)$. We can choose\footnote{While the choice $\RB$ allows for arbitrary volume eigenvalues $\rho$, a subset $\rho \in 2 \mathbb Z + \epsilon$, $\epsilon \in \mathbb R$ is superselected by the action of the Hamiltonian on a state $h^\epsilon_x$.} $\rho$ to be an integer, corresponding to the group U$(1)$, or $\rho \in \mathbb R$, corresponding to $\RB$. It is also possible to introduce a Barbero-Immirzi-like parameter, but we will not do so for notational simplicity, see \cite{BVI} for more details.  
Point holonomies of $P_\alpha$ act by multiplication, while
the conjugate momentum $\alpha$ is smeared over three-dimensional compact regions $R$ and acts as 
\be
	\widehat{\alpha(R)} \ket{\Psi}  = i \sum_{x \in \Sigma} s(R,x)  \frac{\partial}{\partial P_\alpha(x)} \ket{\Psi} \text{,}
\ee
where
\be
	 s(R,x)  :=  \sign (R)  ~ \times ~  \begin{cases} 
										1 &\mbox{if } x  \in R \backslash \partial R \\ 
									1/2 & \mbox{if }  x \in \partial R \\  
									0 &  \mbox{otherwise,}					
								\end{cases} 
\ee
and $\sign(R)$ denotes the orientation of $R$ in $\Sigma$. 
The scalar product is simply given by 
\begin{align}
	\braket{\Psi}{\Phi}_\text{kin} &= \int \prod_{x \in\Sigma} d \mu_{H}(x) \bar{\Psi} \Phi \text{,}
\end{align}
where $d \mu_{H}$ is the normalised Haar measure of U$(1)$ or $\RB$ respectively. 
The additional scalar field is quantised accordingly, as well as the $\beta$ and $\gamma$ variables by including point holonomies of $P_\beta$ and $P_\gamma$. 

We are now in a position to impose the reduction constraints, or more precisely a first class subset thereof. The implementation of the constraints measuring the departure from $q_{ab} \propto \delta_{ab}$ and $P^{ab}  \propto  \delta^{ab}$ is straight forward and given in \cite{BVI}. It simply amounts to implementing $\hat \beta \ket{\Psi} = \hat \gamma \ket{\Psi}= 0$, which restricts the wave functions to have only trivial dependence on point holonomies of $P_\beta$ and $P_\gamma$. The vanishing of their conjugate momenta cannot be imposed simultaneously, as this would amount to imposing strongly second class constraints.
Spatial diffeomorphisms acting on $\alpha$, $P_\alpha$ and $\phi$, $P_\phi$ which are generated by \eqref{eq:DiffRedConstr} are implemented in the standard way following \cite{AshtekarQuantizationOfDiffeomorphism}, see also \cite{BVI}.

It then remains to quantise the Hamiltonian constraint and to investigate its kernel. The terms in the Hamiltonian constraint can be split into three groups:
\begin{enumerate}
	\item {\bf FRW terms:}\\
	These terms give the $k=0$ FRW Hamiltonian upon a classical symmetry reduction. In the quantum theory, these terms will give the LQC Hamiltonian constraint. The relevant terms are
	\be
		 C_\text{FRW, k=0} = \frac{P_\phi^2}{2 \sqrt{q}} - \frac{\left( P^{ab} q_{ab}\right)^2}{3 \sqrt{q}} = \frac{P_\phi^2}{2 \alpha} - \frac{ 3 P_\alpha^2 \alpha}{4} \label{eq:HFRW}
	\ee
	
	  \item {\bf Non-FRW terms without spatial derivatives:}\\
	  These terms account for the trace free parts of $P^{ab}$. They are proportional to reduction constraints and their action thus vanishes on quantum states annihilated by the reduction constraints in a suitable ordering that we choose \cite{BVI}. 
	  
	 \item {\bf Spatial derivatives:}\\
	All other terms contain spatial derivatives. In \cite{BVI}, a finite-difference regularisation on a given underlying graph was employed to define the corresponding quantum operators. The choice of a single-vertex graph then enforces that all such terms have a vanishing action since the vertex was its own neighbour, and therefore a suitable finite-difference regularisation such as $f'(x_{i}) := \frac{f(x_{i+1}) - f(x_{i-1})}{2 \Delta x}$ vanishes. In this paper, we will address this differently, as discussed in the next section.

\end{enumerate}

Our strategy to quantise the Hamiltonian constraint will be driven by purely practical consideration and we do not aim for a unified treatment that addresses questions such as anomaly freedom in a setting beyond the spatially flat, homogeneous and isotropic one. In particular, spatial derivatives will vanish only in expectation values, whereas the FRW part of the Hamiltonian constraint and the other reduction constraints will annihilate physical states. Via this strategy, we can focus on the FRW part.
Following standard LQG quantisation techniques and for a certain choice of lapse, it was shown in \cite{BVI} that the quantisation of \eqref{eq:HFRW} gives the difference equation that one finds in LQC \cite{AshtekarRobustnessOfKey} when acting on a single-vertex state, i.e. a state where the total volume of the universe is encoded in a single node of the underlying graph. The Hilbert space $\mathcal L^2(\RB, d\mu_H)$ at this node is identified with that of LQC and the action of operators thereon directly translates.

\section{Refinement-invariant dynamics} \label{sec:Refinements}

In this section, we will first discuss the dynamics induced on a single vertex, i.e. a single FRW patch, neglecting the spatial derivative terms in the Hamiltonian constraint. We will investigate the resulting dynamics and their invariance under a refinement of the state. Finally, we will discuss a certain highly symmetric choice of physical states in which the expectation value of spatial derivatives vanishes and take a look at standard deviations. 

\subsection{Single patch dynamics}

In order to obtain the dynamics, we quantise \eqref{eq:HFRW} and compute its action on a state satisfying the reduction constraints. For now, we simply neglect spatial derivatives and comment more on them in section \ref{sec:SpatialDer}. It is our aim here to use the exact solution of the quantum dynamics given in \cite{AshtekarRobustnessOfKey}. For this, we rewrite the classical Hamiltonian constraint first as (the reason for this form becomes clear only after a variable transformation \cite{AshtekarLoopQuantumCosmologyBianchi, CraigConsistentProbabilitiesInLoop})
\be
	P_\phi^2 =  \frac{3}{2} \sqrt{|\alpha|} \, P_\alpha \, |\alpha| \, P_\alpha \, \sqrt{|\alpha|} \label{eq:HamConstrSquare}
\ee
and then approximate $P_\alpha$ by $\sin P_\alpha = \frac{e^{i P_\alpha} - e^{- i P_\alpha}}{2i}$. This step is necessary since $P_\alpha$ does not exist as an operator, whereas point-holonomies of $P_\alpha$ do. In particular, this approximation is good as long as the matter energy density is small compared to the Planck density \cite{BVI}, and thus well motivated.

In order to define the final constraint operator, we take square roots on both sides of \eqref{eq:HamConstrSquare} and integrate over $\Sigma$. This corresponds to a deparametrisation w.r.t. to the scalar field $\phi$ as a time variable. The square roots are defined via the spectral theorem, that is we first need to construct the operators corresponding to both sides of \eqref{eq:HamConstrSquare}. 

We want to compute the action of these operators on a state based on a regular cubic graph $\gamma$ with $N^3$ vertices. We construct a cubulation of $\Sigma$ dual to this graph such that the vertices of $\gamma$ are the center points of the cubes. An integral over $\Sigma$ then splits into a sum of integrals over cubes. These integrals are then in turn approximated using the techniques of \cite{ThiemannQSD1, ThiemannQSD5, GieselAQG1, GieselAQG2}, as for example shown in \cite{BVI} in the context of the present variables. 
On each cube $c_i$ of the cubulation, we approximate
\be
	\int_{c_i} d^3x \, f(x) \approx \epsilon^3 f(v_i) \text{,}
\ee
where $v_i$ is the vertex of $\gamma$ in the center of $c_i$ and $\epsilon^3$ the coordinate volume of $c_i$. In particular, the action of the point holonomies, e.g. $f(x) = h_x^\rho$, is at the vertices. Thus, the Hilbert space associated to $\alpha$ and $P_\alpha$ restricted to the graph $\gamma$ is the product of $N^3$ copies of $\mathcal L^2(\RB, d\mu_H)$. 
Since the right-hand side of \eqref{eq:HamConstrSquare} has a square root of a volume operator ordered to the right, it will not create new vertices, but merely act on existing ones. The same is true for the operator corresponding to the left hand side due to $P_\phi$ becoming a derivative w.r.t. $\phi$.
The construction is very similar to that in lattice QFT, with the main conceptual difference that the location of the underlying lattice points is only specified w.r.t. other lattice points due to the diffeomorphism (reduction) constraint \eqref{eq:DiffRedConstr} that we imposed.  

The action of the resulting constraint operator at a single vertex thus reduces to the one familiar from \cite{AshtekarRobustnessOfKey}. The main virtue of the chosen ordering is that the resulting system is exactly soluble and equivalent to the 1+1-dimensional Klein-Gordon equation
\be
	\partial_\phi^2 \Psi(x, \phi) = \partial_x^2 \Psi(x, \phi) =: - \hat \Theta \Psi(x, \phi) 
\ee
after a variable transformation.
It is most instructive to view $\phi$ as an internal time of the system. Physical evolution of the gravitational degrees of freedom in scalar field time $\phi$ is then governed by the operator $\sqrt{\hat \Theta}$ and we can write expectation values of a Schr\"odinger picture operator $\hat{\mathcal{O}}$ at scalar field time $\phi$ as $\vev{\hat{\mathcal{O}}}_{\phi}$.
The main result of \cite{AshtekarRobustnessOfKey} is that the expectation value of the volume $V_i = \int_{c_i} d^3x \, |\alpha|$ as a function of the scalar field is given by
\be
	\twopoint{\hat V_i}_{\phi_i} = V_{\text{min}, i} \cosh \left( \phi_i - \phi_{\text{bounce}, i} \right) \text{,} \label{eq:VevV}
\ee 
where $V_{\text{min}, i}$ and $\phi_{\text{bounce}, i}$ depend on the quantum state at the vertex $v_i$. Since $\hat P_\phi = \sqrt{\hat \Theta}$ generates the evolution, $\vev{\hat  P_\phi}_\phi$ is independent of $\phi$, i.e. $P_\phi$ is a constant of motion. We will therefore mostly drop the $\phi$-subscript form expectation values of $\hat P_\phi$.
Furthermore, the energy density $\rho_i(\phi_i) := \frac{\vev{\hat P_{\phi, i}}^2}{2 \vev{\hat V_i}_{\phi_i}^2}$ is bounded from above by a critical energy density $\rho_\text{crit}$ (at the order of the Planck scale) and reaches its maximum $\rho_{\text{bounce}, i}$ at $\phi_i = \phi_{\text{bounce}, i}$. 
While $\rho_\text{bounce} = \rho_\text{crit}$ is satisfied if one considers the effective classical dynamics, quantum states with finite spread always have $\rho_\text{bounce} < \rho_\text{crit}$.

\subsection{Coarse grained dynamics and state refinements}

In the absence of spatial derivatives, the Hamiltonian constraint operator acts at each vertex $v_i$ separately as discussed in the previous section. Therefore, we will have the same type of solution leading to \eqref{eq:VevV} at each vertex. For the total volume $V$, it follows that 
\be
	\vev{\hat V}_\phi = \sum_{i=1}^{N^3} \vev{\hat V_i}_\phi = V_{\text{min}} \cosh \left( \phi - \phi_{\text{bounce}} \right) \text{,}
\ee
where we used $V_{\text{min}, i}>0$ and (for $a_i>0$)
\be
	\sum_i a_i \cosh (x + b_i) = a \cosh(x+b), ~~ \text{with} ~~ \tanh (b) = \frac{\sum_i a_i \sinh (b_i)}{\sum_i a_i \cosh (b_i)} ~~ \text{and} ~~ a = \frac{\sum_i a_i \cosh (b_i)}{\cosh (b)} \text{.} \label{eq:CoshAddition}
\ee
We draw the important conclusion that the qualitative form of the expectation value of the coarsest observables, the total volume and the total (constant) scalar field momentum $\vev{\hat P_\phi} = \sum_i \vev{\hat P_{\phi, i}}$, is insensitive to the details the underlying state. 
In particular, the final form is insensitive to the number of vertices in the underlying graph. 

The resulting physical picture is that of non-interacting FRW patches whose extensive quantities, the volume and the scalar field momentum, are summed up to obtain the corresponding quantities for the whole universe. The coarse grained trace of the extrinsic curvature as a function of scalar field time can be obtained from the just mentioned Dirac observables by using the Hamiltonian constraint. 
We note that the coarse grained bounce energy density $\rho_\text{bounce}^{\text{c.g.}} := \frac{\vev{\hat P_\phi}^2}{2 V_{\text{min}}^2}$ derived from the coarse grained quantities is maximised
for fixed $\vev{\hat P_{\phi, i}}$ and $V_{\text{min}, i}$ if and only if all the $\phi_{\text{bounce}, i}$ are equal. If furthermore all the $\rho_{\text{bounce},i}$ are equal, we have $\rho_\text{bounce}=\rho_{\text{bounce},i}$.

These considerations suggest a straight forward way to refine a quantum state associated to a single vertex in such a way that the dynamics of the coarse observables remain unchanged. For example, we could evenly subdivide the cube into $N^3$ new cubes. On them, we prescribe $N^3$ quantum states such that their coarse values for $V_\text{min}$, $\phi_\text{bounce}$, and $P_\phi$ match those of the original cube. 
The simplest way to do this would be to evenly split $V_\text{min}$ and $\vev{\hat P_\phi}$ among the $N^3$ cubes and set $\phi_{\text{bounce},i}=\phi_{\text{bounce}}$, which can be interpreted as a rescaling of the coarse FRW patch to arrive at the refined ones.

We stress that in such a refinement step, we have to assume that quantum states with these {\it scaling properties} actually exist. This seems to be a mild assumption, but constructing them explicitly turns out to be difficult, and we will not pursue this here. To give an example, we note that for a specific choice of coherent states \cite{CorichiCoherentSemiclassicalStates}, $\rho_\text{bounce}$ depends on the coherent state parameters $k_0$ and $\sigma$ (and $n=0$) with $\alpha^2 = 12 \pi G$ in their range of validity as 
\be
	\rho_\text{bounce} = \rho_\text{crit} ~ e^{-\alpha^2/\sigma^2}  \frac{1+\frac{\sigma^2}{4 k_0^2}}{1+\frac{\sigma^2+\alpha^2}{4 k_0^2}} < \rho_\text{crit}
\ee
and thus also on the bounce volume. It was therefore concluded in \cite{CorichiOnTheSemiclassical} that those specific coherent states do not satisfy the scaling property that is required for the coarse graining above to leave the dynamics invariant, whereas the general question of the existence of such states was left unanswered. 
The motivation behind \cite{CorichiOnTheSemiclassical}, the independence of the LQC dynamics on the chosen fiducial cell, thus turns out to be closely related to the question of coarse graining discussed here.

We note that the refinement step could in principle be more general, involving in particular unequal $\phi_{\text{bounce}, i}$, as long as all the above requirements are satisfied. However, from the point of view of having a homogeneous and isotropic state, taking equal $\phi_{\text{bounce}, i}$ seems more appropriate. 
This is also motivated by the next subsection, where spatial derivatives are shown to have a vanishing expectation value on a set of highly symmetric states featuring equal $\phi_{\text{bounce}, i}$. On the other hand, having varying $V_{\text{min}, i}$ can be interpreted as having different refinement scales at different lattice sides, which makes sense in particular when considering general underlying graphs and is thus consistent with homogeneity and isotropy. In this case however, spatial derivatives need to be removed by hand.

Should it turn out that there is some fundamental obstacle in constructing quantum states with the necessary scaling property, one could still absorb the mismatch in a running of the Barbero-Immirzi parameter $\gamma$ depending on the quantum state, as $\rho_\text{crit}$ scales as $\gamma^{-3}$. The idea of a running $\gamma$ is not new and has been discussed in \cite{DaumRunningImmirziParameter, BenedettiPerturbativeQuantumGravity, DaumEinsteinCartanGravity, BNI} from different points of view.  

The picture emerging here is very similar to that of ``lattice loop quantum cosmology'' \cite{Wilson-EwingLatticeLoopQuantum}, where an underlying lattice was assumed with one copy of loop quantum cosmology per vertex. The main conceptual difference is that we consider the emerging picture as a quantum state in a quantisation of continuum general relativity, whereas \cite{Wilson-EwingLatticeLoopQuantum} starts with a fixed lattice. The scope of \cite{Wilson-EwingLatticeLoopQuantum} goes beyond the homogeneous and isotropic setting, which may be of interest also here. A similar setting is also discussed in \cite{PawlowskiObservationsOnInterfacing}, focussing on general aspects of an interface between loop quantum gravity and loop quantum cosmology.

\subsection{Spatial derivatives} \label{sec:SpatialDer}

As discussed above, the action of the spatial derivative terms in the Hamiltonian constraint does not vanish any more when going beyond a single vertex. We therefore have two choices: we could simply neglect those terms or we could restrict our wave functions in such a way that spatial derivative terms vanish, at least as expectation values. The latter can be achieved as follows. 

Inspired by recent progress in group field theory using condensate states \cite{GielenHomogeneousCosmologiesAs}, we consider wave functions of the type
\be
	\Psi_{\text{sym}}(h^{\rho_1}_{v_1}, \ldots, h^{\rho_n}_{v_n})  = \psi(h^{\rho}_{v_1}) \psi(h^{\rho}_{v_2})  \ldots \psi( h^{\rho}_{v_n}) \text{,}
\ee
i.e. the wave function is the same at each vertex and the total wave function is a product of the vertex wave functions $\psi$. 
Since the wave function is the same at all spatial points, we are dealing with a heuristic notion of homogeneity and isotropy. 

Consider now a finite difference regularisation of a spatial derivative of the type ${2 \Delta x} f'(x_{i}) := f(x_{i+1}) - f(x_{i-1})$, where $x_{i-1}$, $ x_i$, and $x_{i+1}$ refer to neighbouring vertices along a chosen direction. The corresponding operator will act with $\hat f$ at at $x_{i+1}$ and with $- \hat f$ at $x_{i-1}$. 
If we now compute the expectation value $\braopket{\Psi_\text{sym}}{\hat{f'}}{\Psi_\text{sym}}$, we will obtain twice the same term, however with opposite sign, and thus a vanishing result. A similar argument also works for terms involving multiple spatial derivatives.

We take the vertex state $\psi$ to be a solution to the Hamiltonian constraint (as in \cite{AshtekarRobustnessOfKey, BodendorferAnElementaryIntroduction}) when neglecting spatial derivatives. $\braopket{\Psi_\text{sym}}{\hat{f'}}{\Psi_\text{sym}} = 0$ is preserved by the evolution in scalar field time and the approach of neglecting the spatial derivatives is self-consistent.  
A short computation shows that the variance of $\hat f'$ is determined by that of $\hat f$ at a vertex. 
Let us finish by remarking that a naive finite difference regularisation of spatial derivatives as done here can introduce anomalies in the constraint algebra, which prohibit a straight forward Dirac quantisation. We will not touch this point here further, as our quantisation procedure departs from a pure Dirac quantisation for the spatial derivatives. More discussion on this point can for example be found in \cite{CampigliaUniformDiscretizationsA}.

\subsection{Standard deviations} \label{sec:Variances}

It is instructive to see how the standard deviations $\Delta \hat V_i$ and $\Delta \hat P_{\phi, i}$, where $\Delta \hat {\mathcal O} = \sqrt{\vev{\hat{\mathcal O}^2} -\vev{\hat{\mathcal O}}^2}$, can be chosen to change under the refinement so that the coarse standard deviations remain unchanged if computing in the refined system. We note that both $V_i$ and $P_{\phi, i}$ are integrated densities, meaning that their commutator will naturally also be an integrated density. This gives a somewhat different behaviour as if one of the two would correspond to a scalar, such as $P_\alpha$. 
We again note that since we have not explicitly constructed such states, we only discuss here restrictions put on them by the laws of quantum mechanics. The purpose of this section is to motivate that states with the necessary properties should in principle exist and to highlight some of their properties. 
 
We consider an equal subdivision of the $i$-th vertex into $N$ vertices. The new expectation values at each new vertex $v_{i,j}$ are chosen to be $\vev{ \hat V_{i,j}}_\phi = \frac{1}{N} \vev{ \hat V_i}_\phi$ as well as $\vev{ \hat P_{\phi, i,j}} = \frac{1}{N} \vev{ \hat P_{\phi, i}}$ for $j = 1, \ldots N$. This of course does not determine the refined quantum state completely, in particular not the standard deviations obtained from repeated measurements. It will be our goal to adapt the standard deviations in such a way that they are consistent with the error propagation obtained from independent measurements of all refined constituents.

The Heisenberg uncertainty relation puts a fundamental limit on the standard deviations of two non-commuting operators $\hat A$ and $\hat B$ in a given state: $ \Delta \hat A \cdot \Delta \hat B  \geq  \frac{1}{2} \left| \vev{ \left[  \hat A,  \hat B \right]} \right|$. In our case, we obtain 
\begin{align}
	\left. \Delta \hat V_j \cdot \Delta \hat P_{\phi, j} \right|_{\phi} &\geq \frac{1}{2} \left| \vev{ \left[  \hat V_j,  \hat P_{\phi, j} \right]}_{\phi} \right| \nonumber \\
	 					&= \frac{1}{2} \left| \left. \frac{d}{d \tilde \phi} \right|_{\tilde \phi = 0} \vev{e^{- i \sqrt{\hat \Theta_j} \tilde \phi}  \, \hat V_j \, e^{ i \sqrt{\hat \Theta_j} \tilde \phi}}_{\phi}  \right| \nonumber \\
						&= \frac{1}{2} \left| \left. \frac{d}{d \tilde \phi} \right|_{\tilde \phi = 0} \vev{ \hat V_j }_{\phi+\tilde \phi}  \right| \text{,} \nonumber \\
						&= \frac{1}{2} \left| \left. \frac{d}{d \tilde \phi} \right|_{\tilde \phi = 0} V_{\text{min}, j} \cosh(\phi + \tilde \phi) \right| = \frac{1}{2} V_{\text{min}, j} \left| \sinh(\phi)  \right| \text{.}
\end{align} 
In particular, since $\cosh(\phi) \geq |\sinh(\phi)|$, it is not in violation with the uncertainty relation to have  $\Delta \hat V_j \cdot \Delta \hat P_{\phi, j} = c \vev{V_j}_{\phi}$ for some constant $c \geq 1/2$. This suggests the following: 
we lower the uncertainties in the volume and the scalar field momentum in an equal manner and get $\Delta \hat V_{i,j} = \frac{1}{\sqrt{N}} \Delta \hat V_i$ and  $\Delta { \hat P_{\phi, i,j}} = \frac{1}{\sqrt{N}} \Delta{ \hat P_{\phi, i}}$. It follows that while both standard deviations can be made arbitrarily small if we take $\Delta \hat V_i$ small enough, ratios like $\frac{\Delta \hat V_i }{\vev{\hat V_i}}$ grow arbitrarily under the so defined refinement operation\footnote{Since we lower both $P_\phi$ and $V_i$ equally in the refinement operation, the matter energy density is unchanged and there is no physical inconsistency in going below the Planck {\it length} scale {\it in our minisuperspace approximation}. On the other hand, if one were to operationally define a measurement of a region lower than the Planck length scale, e.g. via a scattering of a photon above Planck energy, one would run into the usual difficulties of an expected black hole formation prohibiting such a measurement due to reaching the Planck energy density.}.  

Let us consider now a measurement of the volume $V_i$ encompassing the $N$ vertices that our original vertex was subdivided into. Since the measurements are uncorrelated, we have $\left(\Delta \hat V_i \right)^2 = \sum_{j=1}^N  \left(\Delta  \hat V_{i,j} \right)^2 = N \left(\frac{1}{\sqrt{N}}  \Delta \hat V_i \right)^2=\left(\Delta \hat V_i \right)^2$. We thus see that also standard deviations can be preserved under the most natural way of refining. The fact that we get arbitrarily large ratios $\frac{\Delta \hat V_i }{\vev{\hat V_i}}$ in the individual cells is not in conflict with this, and thus cannot be taken as a measure of semiclassicality. Rather, only the ratio $\frac{\Delta \hat V }{\vev{\hat V}}$ of the coarse observable should be considered for this question.

\section{Conclusion} \label{sec:Conclusion}

In this paper, we explained how previous results \cite{BVI} of embedding loop quantum cosmology into a full quantum gravity setting can be extended beyond the truncation of the full theory to a single vertex. 
The issue of coarse graining and state refinements was discussed and shown to be closely related to the question of fiducial cell independence in loop quantum cosmology \cite{CorichiOnTheSemiclassical}.
It was shown that the dynamics of the coarsest observables, the total volume of the universe and the scalar field momentum, are unaffected by refinements of the underlying quantum state (with equal $\phi_{\text{bounce}, i}$), i.e. by adding new vertices to the underlying graph and subdividing the total volume and scalar field momentum among them, if a certain scaling condition on the states holds. The graph-preserving regularisation of the Hamiltonian constraint can thus be justified a posteriori by the observation that the final result is independent of the chosen coarseness of the graph. If it will turn out that the scaling condition is impossible to implement, it would be still be possible to absorb the scale dependence of the bounce energy density $\rho_\text{bounce}$ into a running of the Barbero-Immirzi parameter. 
We also have neglected the issue of inverse triad corrections \cite{BojowaldInverseScaleFactor} which appear for different choices of factor ordering and lapse. Since those corrections become large when the volume approaches the Planck volume, their fate is very relevant in the context of refining the underlying quantum state, as one will eventually reach the Planck volume.

The results of this paper thus establish a close connection between LQC-specific questions and the issue of coarse graining in full LQG. While one can argue in LQC that the volume of the universe is large and therefore the scaling property approximately holds \cite{CorichiOnTheSemiclassical} and inverse triad corrections can be neglected, this ceases to be the case in our full theory embedding of LQC once the quantum states are sufficiently refined. Since it is the general expectation that the true cosmological dynamics of LQG should emerge if one computes with very fine fundamental graphs and then coarse grains, the unresolved fate of the scaling property at low volumes turns out to be a highly relevant research questions that should be addressed in more detail. We should warn the reader that in order to extract the cosmological dynamics of LQG, it may also be necessary to consider fundamental quantum states that are symmetric only on larger scales and may fluctuate arbitrarily at small scales. It is unclear to us to which extend this may influence the coarse dynamics. Since full control on such general quantum states seems out of reach at the moment, we think that it is reasonable for now to concentrate on fundamentally symmetric states and to reuse as much as possible of the techniques developed in LQC.

Neglecting spatial derivatives is heuristically motivated by working in a cosmological setting. It was shown that for a certain choice of physical state which mimics a homogeneous and isotropic spacetime, the expectation value of operators obtained from spatial derivatives vanishes. The motivation for considering this type of state comes from recent progress obtained using condensate states in group field theory \cite{GielenHomogeneousCosmologiesAs, OritiEmergentFriedmannDynamics}. Whereas in group field theory the number of vertices is dynamical and large quantum numbers seem to be dynamically suppressed \cite{GielenEmergenceOfA}, we see no such effect here because we restricted ourselves to a graph preserving regularisation and thus excluded this by hand.   
The restriction to a cubic graph was mainly for convenience, in particular to motivate the approximation of neglecting spatial derivatives as discussed in section \ref{sec:SpatialDer}. As long as spatial derivatives are neglected or a similar argument as in section \ref{sec:SpatialDer} is made, our results hold for general graphs and numerous other subdivision schemes along the same lines are possible. 

An important conceptual lesson from our investigation is that {\it if the scaling condition holds for arbitrary rescalings}, the refinement can also be done arbitrarily often, leading in particular to arbitrarily small expectation values of the individual volumes associated to the vertices. 
Such a limit can be considered as a continuum limit for the theory where the underlying graph is chosen arbitrarily fine. This limit is not in contradiction with the discrete spectrum of the volume operator, which is invariant under refinements, due to superpositions of different volume eigenstates leading to the arbitrarily small expectation values. Also, the relevant quantity determining the onset of large quantum gravity effects, the energy density, is invariant under the refinements. Thus, quantum gravity effects such as a big bounce replacing the big bang close to the critical (Planck) energy density \cite{AshtekarQuantumNatureOf} or modifications in dispersion relations due to a deformation of the hypersurface deformation algebra depending on the energy density \cite{BojowaldDeformedGeneralRelativity, MielczarekLoopDeformedPoincare, RoncoOnTheUV, AmelinoSpacetimeNoncommutativityRegime, BodendorferAnElementaryIntroduction} are surviving this continuum limit\footnote{This continuum limit should not be confused with an alternative limit discussed in \cite{CorichiOnAContinuum}, which sends the volume gap to zero, i.e. the lattice size of the lattice on which the Hamiltonian constraint acts in the $v$-representation.}.
On the other hand, our result casts doubt on physical predictions which rely on a fixed underlying graph to provide a lattice-like structure at the Planck scale, such as Lorentz violations at low (and often experimentally excluded \cite{LiberatiTestsOfLorentz}) orders in $E/E_{\text{Planck}}$. A lattice-like structure would still be present at finite refinement in our case, but the effective lattice spacing (computed from expectation values of the cell volumes) can be made arbitrarily small.

For future work, it will be interesting to extend the present results to full theory embeddings of Bianchi I cosmology \cite{BIII} as well as spherical symmetry \cite{BLSI, BVI}. Also, the current way of strongly imposing the reduction constraints $\beta=0$ and $\gamma=0$ leaves their canonically conjugate variables with maximal uncertainty. It may be more satisfying to impose the $\beta=0$, $P_\beta = 0$, $\gamma = 0$, and $P_\gamma=0$ weakly using a similar construction as in \cite{ThiemannComplexifierCoherentStates}, see also \cite{BeetleDiffeomorphismInvariantCosmological} for related work.

\section*{Acknowledgements}
This work was supported  by the Polish National Science Centre grant No. 2012/05/E/ST2/03308, an FNP START fellowship, and during final improvements by a Feodor Lynen Return Fellowship of the Alexander von Humboldt Foundation. The author thanks Daniele Oriti and Edward Wilson-Ewing for enlightening discussions on the topic of coarse graining and the interpretation of condensate states in group field theory, as well as Parampreet Singh for an email exchange about the LQC dynamics.


\end{document}